\documentstyle[12pt]{article}
\title{A Dipole interpretation of
the $\nu=1/2$ state}
\author{V.Pasquier\\
CEA/Saclay, Service de Physique Th\'eorique\\
F-91191 Gif-sur-Yvette Cedex, FRANCE\\
F.D.M Haldane\\Department of Physics, Princeton University, \\Princeton, New Jersey 08544}

\begin{document}
\maketitle
\abstract{
We consider the problem of Bosonic particles interacting repulsively
in a strong magnetic field at the filling factor $\nu=1$.
We project the system in the Lowest Landau Level 
and set up a formalism to map
the dynamics into an interacting Fermion system. 
Within a mean field approximation we find that
the composite Fermions behave as a gas of neutral dipoles and
we expect that the low energy limit also
describes the physical $\nu=1/2$ Fermionic state.}

\vfill\eject

\section{Introduction}

There has been recently a renewed interest in the 
quantum Hall effect when the filling factor is a
fraction with an even denominator.
Willets and his collaborators\cite{WIL} have observed an anomalous behavior in
the surface acoustic wave propagations near $\nu=1/2$ and $\nu=1/4$.
A remarkable outcome of their experiments
is that they probe a longitudinal conductivity $\sigma_{xx}(q,\omega)$
increasing linearly with the wave vector $q$.
Halperin, Lee and Read \cite{HLR} have suggested that
the system exhibits
a Fermi liquid behavior at this particular value. 
They have developed a formalism based on the Chern-Simon
theory 
which provides an explanation for the experimental observations.
Subsequently, several studies have developed and improved
the predictions of the
Chern Simon theory \cite{HLR}  
\cite{CS1} \cite{CS2} \cite{CS3} \cite{CS4}. 
Another approach followed by Rezayi and Read
\cite{REZ} and  Haldane et al \cite{HAL1} consists in
obtaining trial wave functions to study
numerically the properties of the system at this filling factor.
In these studies the
cyclotron frequency is supposed to be sufficiently large so that the only
relevant excitations are confined to the lowest Landau level.
The trial wave functions can be compared to the exact ground state
and the overlap between the two turns out to be extremely good \cite{REZ}.
In these studies
the effective mass $m^*$ which defines the Fermi velocity
is generated dynamically by the
interactions. 
This paper introduces a microscopic model 
closely related to these trial wave functions.
We have considered toy model of Bosonic particles
interacting repulsively in a magnetic field at a filling factor
$\nu=1$. 
Although it may at first look different,
the problem of 
formation of a Fermi sea is essentially the same as in
the $\nu=1/2$ case.
If one applies the analyses of the composite Fermions \cite{JAI}
or the Chern Simon approach to such a system, one is led
to the same picture of Fermi sea formation as in the $\nu=1/2$
case. 
Halperin Lee and Read use the composite Fermions arguments to motivate
the formation of a Fermi sea in the $\nu=1/2$ case.
They attach two magnetic fluxes to an electron in order to
cancel the magnetic field seen by the electron in the mean field
approximation. This flux attachment does not modify the
statistics of the electrons 
and if one ignores fluctuations one is led to a system of
spinless Fermions in a zero magnetic field.
In the case of Bosons at $\nu=1$, we can proceed similarly
by attaching one flux unit to each particle
so as to cancel the exterior magnetic field.
In this process the statistics is changed from Bosons to Fermions
and a Fermi liquid is expected to form.

Read has interpreted the fluxes attached to
the electron as physical vortices bound to it \cite{READ}(see also 
\cite{READ1} and \cite{BAS}).
We claim that his proposal differs from the 
mean field interpretation for the following reason.
The mean field treats the composite electron as
a charged particle which couples to the electro-magnetic field.
The vortices carry a charge equal to minus one half of that of the electron
so that the bound state is
as a neutral particle which propagates in a constant
charge background.
In this case 
the response to an external field depends on the dipole
structure of the composite object.
We are led to this picture in
the $\nu=1$ case. The main simplification is
that there is a single vortex coupled to the Boson, this vortex
is a Fermion carrying the opposite charge and we can
use a  second quantized formalism to analyses the model.
The bound state is then a dipole 
\cite{GOR} \cite{LER} \cite{KAL}.

Our approach is mainly 
motivated by the trial wave functions of \cite{REZ} \cite{HAL1}.
If the particles were distinguishable
the Laughlin wave function \cite{LAU}
would be the best ground state
but it
gives the particles the wrong statistics.
One corrects for this by multiplying
the trial wave functions
by a Slater
determinant of plane waves and project
the product into the lowest Landau level (LLL).
The effect of the projection is to replace the coordinate 
in the plane waves by operators\cite{REZ}
which
displace the particles 
from their original position.
Here we advocate that the 
charge fluctuations induced by this displacement
are the fundamental excitations.

The next section presents the microscopic model
and analyze its phenomenological consequences.

\section{The Microscopic Model}
\subsection{motivation}

Consider $N$ particles of identical charge interacting with a 
repulsive force in a domain of area $\Omega$
thread by a magnetic field $B$. $B$ is
chosen so that the flux per unit area
is equal to one. 
The magnetic length
$l=\sqrt{\hbar c/eB}$ is such that $\Omega=2\pi l^2 N$.
We assume that the cyclotron frequency 
is large compared to the 
interaction so that the dynamic can be restricted to the
Lowest Landau Level.
The one body Hamiltonian
has N degenerate eigenstates, thus
in the case where the particles are Fermions the
only accessible state
is given by the slater determinant of the one body wave functions.
Suppose the particles are divided into two sets which differ only
by
the statistics they obey.
The first set contains $N_1$ Fermions and the second set contains $N_2$
Bosons the sum $N_1+N_2=N$ being kept fixed so that the filling factor
remains equal to one  and the interaction are the same between
all the particles.
The simplest case consists of
$1$ Boson interacting with
$N-1$ Fermions. By performing a particle hole transformation
on the Fermions we can equivalently regard this as
a Boson interacting with a hole,
a problem
studied by Kallin and Halperin 
\cite{KAL}. 

In the Landau gauge one body Hamiltonian is proportional to:
\begin{equation}
H=p_{x}^2+(p_{y}-x/l^2)^2
\label{HAM4}
\end{equation}
This Hamiltonian commutes with the two guiding center coordinates
$R_y=l^2p_x-y$ and $R_x=l^2p_y$ which do not 
commute with each other $[R_x,R_y]=il^2$.
In the particle hole case
the Hamiltonian of the 
pair is the sum of two one body Hamiltonians where we
change the sign of the the potential 
vector for the hole.
Since the particles have exactly opposite charges, the guiding
center coordinates of the pair
$R_{y_1}+R_{y_2}=l^2p_{x_1}+l^2p_{x_2}-(y_1-y_2)=l^2p_x$  and 
$R_{x_1}+R_{x_2}=l^2p_{y_1}+l^2p_{y_2}=l^2p_y$ commute with each other
and can be diagonalized simultaneously with $H$.
The wave function which diagonalizes
this generalized momentum 
describes a  dipole propagating freely with its
dipole vector 
$l^2\left(p_y, -p_x \right)$
perpendicular to the momentum
$\left(p_x, p_y\right)$.
The potential interaction 
commutes with the momentum $p_x,p_y$ 
so that these wave functions 
are
eigenstates of the 
total projected Hamiltonian. 
When the interaction between the particles is repulsive
these wave functions describe bound
states of size comparable to the magnetic length with a mass of
the order of $V(l)$.

In the general case, if $N_2/N$ is small compared to 1, it is legitimate to
subdivide the particles and the holes into pairs so as to
include the interaction in each pair in the one body Hamiltonian
$H_0$ and treat 
residual
interaction between the different pairs as a perturbation.
When this ratio is equal to
one ($N=N_2$) it is more 
difficult to argue that the low density approximation is valid.
Nevertheless, because the bound states are Fermions
and if we assume that only the quasiparticles at the Fermi surface participate to the dynamics,
this approximation still makes sense.
In the next  section we set up the formalism based on this general philosophy 
to map the Bosonic problem into a Fermionic one.

\subsection{General Formalism}

We consider the case of a
finite geometry such that
the degeneracy of the LLL is equal to the number of bosons $N$.
A basis of LLL orbitals is indexed by $i, 1\le i\le N$.
To each orbital we associate the canonical
Bosonic creation operator $a^+_i$
which creates a state in this orbital.
The Hilbert state of the $\nu=1$ Bosons is generated by the states where
$N$ creation operators act upon the vacuum state $|0>$ defined by
$a_i|0>=0$. 


One can map the Bosonic space into a 
subspace of a Fermionic
space proceeding as follows:
To the $a^+_i,a_i$ we adjoin a set of canonical
Fermionic operators $f^+_i,f_i$ also labeled by the LLL orbitals
and consider the vacuum $|\Omega>$ obtained by filling
the Fermionic orbitals $f^+_i|\Omega >=
b_i|\Omega >=0$. The Bosonic Hilbert space is recovered upon acting
on $|\Omega>$ with $N$ pair creation operators $b^+_i f_j$.
The idea 
(called the method of images in other contexts \cite{RIP})
is to substitute a creation operators
$\chi^+_{ij}$
for the pair $b^+_i f_j$. 
We thus consider a set of operators defined by:
\begin{eqnarray} 
&&\{\chi_{ij},\chi_{kl}\}=
\{\chi^+_{ij},\chi^+_{kl}\}=0
\nonumber \\
&&\{\chi^+_{ij},\chi_{kl}\}=\delta_{il} \delta_{jk}
\label{FERMCO}
\end{eqnarray}
The Fermionic Hilbert space is obtained upon acting with
$N$ $\chi^+_{ij}$ on the vacuum $|\Omega'>$
annihilated by the $\chi_{ij}$.
This description of the original Bosonic 
space is still overcomplete 
since the pairing between Bosons and Fermions is arbitrary
in the definition of the pairs
$\chi_{ij}$.
To recover the physical space, we must project 
the Hilbert space generated by the $\chi^+_{ij}$ onto
the sub-space  antisymmetric under the
permutations of the Fermionic indices $j$:
\begin{eqnarray} 
b^+_{i_1}...b^+_{i_n}|\Omega>=
1/{N!}\sum_{p\in S_N} (-)^p \chi^+_{i_1{p_1}}... 
\chi^+_{i_N{p_N}}|\Omega '> 
\label{IMA}
\end{eqnarray}

Next we identify the observables
in both representations as follow:
\begin{eqnarray} 
&&\pmatrix{\rho^b_{ij}&A^+_{il}\cr
        A_{kj}&\rho^f_{kl}\cr}
=
\pmatrix{b^+_ib_j&b^+_if_l\cr
        f^+_kb_j&f^+_kf_l\cr}
=
\nonumber \\
&&\pmatrix{(\chi^+\chi)_{ij}&(\chi^+\sqrt{1+:\chi\chi^+:})_{il}\cr
        (\sqrt{1+:\chi\chi^+:}\chi)_{kj}&(1+:\chi\chi^+:)_{kl}\cr}
\label{HOLS}
\end{eqnarray}
Where the normal ordering refers to the vacuum. 

To establish (\ref{HOLS}), we must verify that both sets
of operators obey the same $U(N|N)$ algebra and that the representations
obtained by acting with them upon the respective vacua are equivalent.
It is straightforward to verify that
the diagonal blocks $\rho^b,\rho^f$ 
obey the same commutation
relations. It is less obvious that the
relation:
\begin{eqnarray} 
\{A^+_{kj},A_{il}\}= 
\delta_{ji}\rho^f_{kl}
+ \delta_{kl}\rho^b_{ij}
\end{eqnarray}
are satisfied in the $\chi$ representation
and can be shown along the lines of
\cite{PAPA}
\footnote{These authors consider the more general case
where the particles carry a flavor index ($b_j\to b_j^a)$ taking $n$ values
which is summed in the upper matrix of (\ref{HOLS}).
In this case $1$ must be replaced by $n$ in the lower matrix of (\ref{HOLS}).}.
The operators $A^+_{il}$ in the upper right corner 
are the pair creation operators used to generate the Hilbert space
upon acting on the vacuum. In what follow, we shall 
keep only the first term in the expansion of the square root and
simply replace them by $\chi^+_{il}$.
This amounts to disregard the projection in (\ref{IMA}).

The respective vacua $|\Omega> ,
|\Omega'>$  are both annihilated by the down left block matrices
$A_{kj}$, thus defining highest weight representations of
$U(N|N)$. 
The equivalence of the representations 
follows from action of the diagonal blocks $\rho^{b,f}$
on the vacuum:
\begin{eqnarray} 
\rho^b_{ij}|\Omega>=0,\ \ \ 
\rho^f_{kl}|\Omega>=\delta_{kl}|\Omega>
\end{eqnarray}

The original Boson dynamics can be expressed
in terms of the density operators $\rho^b_{ij}$ which obey the $U(N)$
algebra. To describe the dynamics 
one possibility would be to use the expression of these
operators terms of $\chi$. Here, we treat
the Fermions as real particles which see the external field 
in the same way as the Bosons. This amounts to replace the
density operators by 
\begin{eqnarray} 
\rho_{ij}=:\rho^b_{ij}+\rho^f_{ij}:
=:\{\chi^+,
\chi\}_{ij}:
\label{RHO}
\end{eqnarray}
Since there is no Fermion when $\nu=1$ 
$\rho^f_{ij}$ is essentially equal to zero. 
This modification nevertheless affects the way we approximate the system.
In particular, 
the dynamics is now well defined 
when the number of $\chi^+$ operators which act on the vacuum is not equal to $N$
and this allows us
to vary the density of pairs arbitrarily.

To recover the real space description
let us for concreteness consider the case of a rectangular box of size
$L_x,L_y$ with $L_xL_y=2\pi l^2 N$. 
We set $z=(x+iy)/L_y$ and $\tau=L_x/L_y$.
In these notations the LLL orbitals wave functions are given by:
\begin{eqnarray} 
<\vec x|j>={1\over \sqrt{\pi L_y}}e^{-\pi x^2/\tau}\theta_j(z)
\end{eqnarray}
where $1\le j\le N$ and $\theta_j$ is the theta function defined as:
\begin{equation}
\theta _{j}(z,\tau)=\sum_{n \in Z} 
\exp(-\pi (j+nN)^2 \tau + 2\pi  (j+Nn)z)
\end{equation}
Except for a common factor 
these wave functions depend on $x,y$ only through the variable $z$
and a family of coherent states $|z>$ can be defined \cite{KLAU}
such that $<z|i>=<\vec x|i>$.
Suppose that the LLL particles interact with a scalar potential $V(\vec x-\vec y).$
After projection  the Hamiltonian takes the form:
\begin{eqnarray} 
H=1/2\int \rho(\vec x)\  V(\vec x-\vec y)\  \rho(\vec y)\  d^2x\ d^2y
\label{USER}
\end{eqnarray}
where  $\rho(\vec x)$ is  the projected density
operator 
\begin{eqnarray} 
\rho(\vec x)=<z|\hat \rho|z>=\sum_{i,j} <\vec x|i>\rho_{ij}<j|\vec x>
\end{eqnarray}

The projection relates a field $\rho(\vec x)$ to a
matrix $\hat\rho$ and more generiquely,
the transformation which associates the function $\rho(x)=<z|\hat \rho|z>$ 
to the
matrix $\rho_{ij}$ is called its P-symbol in
\cite{KLAU}.
Since 
translations act naturally on $\rho(x)$
and there are $N^2$ matrix elements
\footnote{since there are only $N^2$ Fourier modes
the system is in fact defined on a square lattice
with a lattice cut-off $a$ equal to $\sqrt{2 \pi}l/{\sqrt{N}}$.
A difficulty is that we have to deal with three length scales
$a<<l<<\xi$ where $\xi$ is the physically relevant scale.}
we can decompose
$\rho(x)$ onto $N^2$ plane waves $\rho(x)=\sum_{\vec k}e^{ikx}\rho_{k}$
with $k_i=2\pi n_i/L$, $0\le n_1,n_2 \le N-1$.
By the inverse transformation, $\rho(x)=
<z|\hat \rho|z>$ where
$\hat \rho= 2\pi l^2 \sum_k  e^{k^2l^2/4} \hat e^k\rho_k$
and the matrices  $\hat e^k$ obey the 
magnetic translation algebra (\cite{GIR}):
\begin{equation}
\hat e^k\hat e^q=e^{il^2(k\times q)/2}\hat e^{k+q}
\end{equation}
This defines a matrix product on functions which we
denote by 
$\star$ to distinguish it from the ordinary product. 

Let $\Psi(x)=\sqrt{2\pi l^2}<z|\hat \chi|z>$ denote 
the space dependent field
associated to the matrix field $\chi_{ij}$.
The commutation relations (\ref{FERMCO}) imply the
following decomposition for
$\Psi^+(x)$:
\begin{eqnarray} 
\Psi^+(x)=1/L^2 \sum_k e^{-k^2/4}e^{ikx}c^+_k
\end{eqnarray}
where $c^+_k,c_k$ are canonical Fermionic operators.

In these notations the density (\ref{RHO}) is given by:
\begin{eqnarray} 
\rho(\vec x)={1 \over 2\pi l^2}:\{\Psi^+\star,\Psi\}(x):
\approx l\vec\nabla \times \Psi^+{il\vec\nabla }\Psi(x)
\label{HAMM}
\end{eqnarray}
The anticommutator originates from the fact that we add the two contributions
$\rho^b$ and
$\rho^f$ treating the pairs as composite particles.
As a result,
the dominant term in a gradient 
expansion is the right-hand 
side of this equality.

The Hamiltonian   (\ref{USER})
can be expressed in terms of
these operators and the most
relevant contributions around a Fermi surface is given by:
\begin{eqnarray}
&&H=
\int d^2x\ \Psi^+ (-\Delta /
{2m^{\ast}}+\mu)
\Psi (\vec x)\ +\int d^2x\ \int d^2 y
\nonumber \\
&& (\Psi^+ il\vec \nabla \Psi(\vec x) \times l\vec \nabla_x )
(\Psi^+ il\vec \nabla \Psi(\vec y) \times l\vec \nabla_y )
V(\vec x-\vec y)
\label{HAMI}
\end{eqnarray}
where
the effective mass $m^*$ is of the order of magnitude of $V(l)$.
Although our derivation is  only valid for the $\nu=1$ case, it is 
tempting to assume that
the low energy limit of this Hamiltonian 
also describes the physical situation $\nu=1/2$. In this case,
the chemical potential $\mu$ must be adjusted so that the density is
equal to $\nu /{2\pi l^2}$ which imply that
the Fermi momentum $k_F=\sqrt{2\nu}/l$. 
Recently,
Shankar and Murthy \cite{SHAN} have derived the same Hamiltonian 
in the $\nu=1/2$ case using
a different aproach.


The  interaction has no Gallilean invariance
which is not surprising 
in the presence of a magnetic field.
If $V(r)$ behaves as $ r^{-1}$ at large distance,
the induced dipole potential
behaves as $r^{-3}$ and we expect no infrared singularity.
The system becomes essentially equivalent to a Fermi
liquid with short range interaction.

Consider now the linear response to a scalar field $\Phi(\vec x,t)$.
In the long wavelength limit
the interacting Hamiltonian is given by:
\begin{eqnarray}
H_i=\int d^2x\  l\vec\nabla \Phi(\vec x,t)\times \Psi^+{il\vec\nabla }
\Psi(\vec x)
\label{LINE}
\end{eqnarray}

As a consistency check we can couple the system
to a constant electric field $\Phi(\vec x)=e\vec E.\vec x$.
In this case the interacting Hamiltonian $H_i=e\vec E\times \vec K$ 
where $\vec K$ is the total momentum. 
Its only effect is to give an additional speed $e|E|$
in the direction perpendicular to $\vec E$ 
to each quasiparticle.
Thus one recovers the value
of the transverse conductivity $\sigma_{xy}=\nu e^2/{2\pi\hbar }$. More precisely, the
vacuum is charged and responsible for the current  while the quasiparticles
are neutral and carried by the the vacuum.

Using the transport equation in presence of the interaction
(\ref{LINE}), one obtains the static response
function and the dynamical form factor:
\begin{eqnarray}
\chi(\vec q,0)&=&-(l^2qk_f)^2 \nu (0)/{2(1+F_1)}
\nonumber \\
S(\vec q,\omega)&=&
S^0(\vec q,\omega)
(l^2k_f q/{1+F_1})^2
\end{eqnarray} 
$\nu(0)=m^*\Omega/{2\pi}$ is the density of states on the Fermi surface.
$S^0(\vec q,\omega)$ is the free fermion form factor and $F_1$ is the
first
Landau parameter ($F_0$ is not relevant in this theory).
The essential difference with the Fermi liquid results are the factors
proportional to $(lq)^2$ which damp the effect of the external field at low $q$
and originate from the fact that dipoles couple weakly
to an external potential.


\subsection{Conclusion}

We have introduced a microscopic model to analyze the problem
of Bosonic particles in a strong magnetic field at $\nu=1$.
We have generalized the model so as
vary the Fermi momentum $p_f$ and
to study it in a mean field approximation.

The present model gives a description in agreement with
the dipole picture introduced by N.Read \cite{READ}.
The main conclusion of our study is that the system behaves essentially
as a gas of Fermionic dipoles with a dipole vector perpendicular to their
momentum.
As a result, the interactions are screened and the gas behaves in
many respect as a neutral Fermi liquid (the same conclusion
is reached in \cite{BAS}).
The main consequence is that the linear response quantities
get renormalized by a factor $(lq)^2$ at low momentum
transfer $q$.
The Landau theory
only relies on the hypotheses that the
quasiparticles are dipoles 
and should therefore also be valid in the $\nu=1/2$ case.

The model differs from the Chern-Simon theory
by the fact that the dynamics is projected into the
Lowest Landau Level and
the effective mass depends only on
the interactions.
This model does not seem to predict 
a divergence of the effective mass.

\bigskip
{\bf Acknowledgments}

We are thankful to
G.Baskaran,
Kun Yang,
S.Nonnenmacher and N.Read
for helpful discussions.
We are also thankful to the Institute for Theoritical Physics, UCSB,
for hospitality while this work was completed.

\vfill\eject
\end{document}